\documentclass[11pt,draftclsnofoot,onecolumn]{IEEEtran}
\usepackage{epsfig,epsf,latexsym,amsfonts}
\usepackage{amsmath,amssymb,verbatim,mathrsfs,graphicx,bm,multicol}
\input{amssym.def}
\input{amssym.tex}

\raggedcolumns

\usepackage{multirow}
\usepackage{slashbox}
\usepackage{hhline}

\newcommand{\ra}{\rightarrow}
\newcommand{\beq}{\begin{equation}}
\newcommand{\eeq}{\end{equation}}

\newcommand{\mc}{\mathcal}
\newcommand{\mb}{\mathbb}

\newcommand{\tl}{\tilde}
\newcommand{\h}{\hat}

\newcommand{\e}{\epsilon}

\newcommand{\Ae}{A_\epsilon^{(n)}}
\newcommand{\Aetl}{A_{\tilde{\epsilon}}^{(n)}}

\newtheorem{thm}{Theorem}
\newtheorem{dfn}{Definition}
\newtheorem{rem}{Remark}
\newtheorem{lem}{Lemma}

\newcommand{\beqn}{\begin{eqnarray}}
\newcommand{\eeqn}{\end{eqnarray}}

\allowdisplaybreaks 

\begin{document}

\title{Distributed Source Coding with One Distortion Criterion and Correlated Messages}

\author{Suhan Choi}

\maketitle

\begin{abstract}
In this paper, distributed (or multiterminal) source coding with one distortion criterion and correlated messages is considered. This problem can be also called ``Berger-Yeung problem with correlated messages''. It corresponds to the source coding part of the graph-based framework for transmission of a pair of correlated sources over the multiple-access channel (MAC) where one is lossless and the other is lossy. As a result, the achievable rate-distortion region for this problem is provided. It is an information-theoretic characterization of the rate of exponential growth (as a function of the number of source samples) of the size of the bipartite graphs which can represent a pair of correlated sources with satisfying one distortion criterion. A rigorous proof of the achievability and the converse part is given. It is also shown that there exists functional duality between Berger-Yeung problem with correlated messages and semi-deterministic broadcast channel with correlated messages. This means that the optimal encoder-decoder mappings for one problem become the optimal decoder-encoder mappings for the dual problem. In the duality setup, the correlation structure of the messages in the two dual problems, source distortion measure and channel cost measure are also specified.
\end{abstract}


\section{Introduction}
Consider a many-to-one communication scenario where many transmitters want to send correlated input sources reliably and simultaneously to one joint receiver over the channel. This scenario has many practical examples such as sensor networks and wireless cellular networks. For this transmission scenario, there have been two approaches. One is the joint source-channel coding  \cite{cover-elgamal-salehi80} and the other is the separate source-channel coding. The latter, which is based on Shannon's separation theorem in point-to-point communication, is divided into two multiterminal coding modules: the distributed source coding (DSC) \cite{slep-wolf1,berger,berger-yeung89} and the multiple-access channel (MAC) coding \cite{ahlswede71,liao}.

In the DSC, many source encoders want to represent their correlated input sources into messages simultaneously and one joint source decoder reconstructs the original sources from the received messages (without channel errors) with satisfying certain distortion criterion. In the MAC coding, many channel encoders wish to send their input messages reliably and simultaneously over the channel, and one joint channel decoder decodes the original messages from the channel output. Here, the messages are used as a discrete interface between source coding and channel coding modules.

Before the recent introduction of graph-based framework \cite{pradhan-choi-ram07}, only \emph{independent} messages are considered in this separation approach. However, it has been generally known that even though this separation approach with independent messages gives convenient modularity, it is generally not optimal in multiterminal communications scenarios \cite{cover-elgamal-salehi80}. In the graph-based separation approach \cite{pradhan-choi-ram07}, \emph{correlated} messages, which can be associated with bipartite graphs, are used as a discrete interface between source coding and channel coding modules. This approach enables us to maintain Shannon-style modularity, and to minimize the performance loss as compared to the optimal joint source-channel coding \cite{cover-elgamal-salehi80}.

More specifically, in the source coding module, the correlated sources are encoded (or represented) distributively into correlated messages. And these correlated messages are then encoded and reliably sent over the channel in the channel coding module. Here, the correlated messages correspond to edges in the graph. This graph-based framework was also applied to the broadcast channel \cite{choi-pradhan08}.

For DSC with a pair of correlated sources (and independent messages), there have been three different problems in terms of distortion criterion:
\begin{itemize}
  \item Lossless DSC (Slepian-Wolf problem \cite{slep-wolf1}) where both sources are lossless,
  \item Lossy DSC (studied by Berger \cite{berger} and Tung \cite{tung78}) where both sources are lossy, and
  \item DSC with one distortion criterion (Berger-Yeung problem \cite{berger-yeung89} where one source is lossless and the other is lossy.
\end{itemize}

Note that Pradhan \emph{et al.} \cite{pradhan-choi-ram07} considered only \emph{lossless} DSC with correlated messages, presenting the achievable rate region for this problem. For the lossy DSC with correlated messages, Choi \cite{choi09} provided an inner bound to the achievable rate-distortion region. Through the result of \cite{choi09}, the graph-based separation framework in \cite{pradhan-choi-ram07} can be extended to the transmission of analog correlated sources. This is because the same channel coding module in \cite{pradhan-choi-ram07} can be used to send the correlated messages which are the outputs of both lossless and lossy source encoders.

In this paper, we study a DSC problem with one distortion criterion and correlated messages. This problem can be also called ``Berger-Yeung problem (BYP) with correlated messages''. In this problem, two non-communicating encoders represent a pair of correlated sources into correlated messages and a joint decoder reconstruct the original sources, where the reconstruction of one source is lossless and that of the other is lossy with certain distortion criterion. In other words, this problem corresponds to the source coding part of the graph-based framework \cite{pradhan-choi-ram07} for transmission of a pair of correlated sources over the MAC, where one is lossless and the other is lossy.

As a result, we provide the achievable rate-distortion region for this problem, which is an information-theoretic characterization of the rate of exponential growth (as a function of the number of source samples) of the size of the bipartite graphs which can represent a pair of correlated sources with satisfying one distortion criterion for the transmission over the MAC. We present both the achievability and the converse part of the proof.
This indicates that a pair of correlated sources can be reliably represented into a nearly semi-regular bipartite graph even when one source is lossless and the other is lossy.

It is also illustrated that a given pair of correlated sources can be efficiently represented into many different nearly semi-regular bipartite graphs without increasing redundancy.

Therefore, based on the merged results of this paper, \cite{pradhan-choi-ram07} and \cite{choi09}, it can be concluded that, for transmission of any (both discrete and continuous) set of correlated sources over the MAC, graphs can be used as discrete interface between source coding and channel coding modules.

We also examine functional duality between our problem, Berger-Yeung problem with correlated messages, and semi-deterministic broadcast channel (SBC) with correlated messages \cite{choi-pradhan08}. Similar researches have been presented in the literature. In \cite{stankovic-cheng-xiong06}, functional duality between Berger-Yeung problem and semi-deterministic broadcast channel was studied. Functional duality between distributed source coding and broadcast channel coding problems is also provided in \cite{pradhan-ram06}. However, it should be noted that previous studies \cite{stankovic-cheng-xiong06} and \cite{pradhan-ram06} considered functional dualities in the case of \emph{independent} messages only. Accordingly, it is natural to ask whether this duality holds for \emph{correlated} messages or not.

Consequently, we show that under certain conditions, for a given BYP with correlated messages, a functional dual SBC with correlated messages can be obtained, and vice versa. This means that the optimal encoder-decoder mappings for one problem become the optimal decoder-encoder mappings for the dual problem. We also specify the correlation structure of the messages in the two dual problems and source distortion measure and channel cost measure for this duality.

The rest of this paper is organized as follows. We first give some preliminaries including the definition of bipartite graphs and the concept of correlated messages in Section \ref{sec:prelim}. In Section \ref{sec:summary}, we formulate the problem and present the achievable rate-distortion region, which is one of the main results of this paper. Thereafter, the proof of the theorem will be provided in Section \ref{sec:thm-proof}. Then, we discuss the representation of a correlated sources into many different graphs in Section \ref{sec:different}. In Section \ref{sec:duality}, we examine the functional duality between BYP with correlated messages and SBC with correlated messages. Finally, Section \ref{sec:conclusion} provides concluding remarks.

\section{Preliminaries}\label{sec:prelim}

Before we discuss the main problem, let us first define a
bipartite graph and related mathematical terms which will be used
in our discussion.
\begin{dfn}
\begin{itemize}
\item A \emph{bipartite graph} $G$ is defined as an ordered tuple
$G=(V_1(G),V_2(G), E(G))$ where $V_1(G)$ and $V_2(G)$ denote the
    first and the second vertex sets of $G$, respectively,  and
  $E(G)$ denotes the edge set of $G$. i.e., $E(G) \subseteq V_1(G) \times V_2(G)$.
\item    If $E(G) = V_1(G) \times
  V_2(G)$, then $G$ is said to be \emph{complete}.
\item The \emph{degree} of a
vertex $u \in V_1(G)$ in a graph $G$, $\mathrm{deg}_{G,1}(u)$, is the number of vertices in
$V_2(G)$ that are connected to $u$. Similarly $\mathrm{deg}_{G,2}(v)$
is defined for all $v \in V_2(G)$.
\end{itemize}
\end{dfn}
Since we use a special type of bipartite graphs, we define those bipartite graphs as follows.
\begin{dfn}\label{def:graph-5-parameters}
 A bipartite graph $G$ is called \emph{nearly semi-regular} with parameters $(\Delta_1$, $\Delta_2$, $\Delta_1'$,
$\Delta_2'$, $\mu)$ for $\mu > 1$, denoted by $G(\Delta_1$, $\Delta_2$, $\Delta_1'$,
$\Delta_2'$, $\mu)$, if it satisfies:
\begin{itemize}
\item $|V_i(G)|=\Delta_i $ for $i=1, 2$,
\item $\forall u \in V_1(G)$, $\Delta_2' \mu^{-1} \leq
  \mathrm{deg}_{G,1}(u) \leq \Delta_2' \mu$,
\item $\forall v \in V_2(G)$, $\Delta_1' \mu^{-1}  \leq \mathrm{deg}_{G,2}(v)  \leq \Delta_1' \mu$
\end{itemize}
where $|A|$ denotes the cardinality of a set $A$.
\end{dfn}

Note that these nearly semi-regular graphs have slackness parameter $\mu$ for the degrees of the vertices. These nearly semi-regular graphs become semi-regular if $\mu=1$.

In our problem, we also consider a pair of \emph{correlated} messages $(W_1,W_2)$ such that $(W_1,W_2) \in \mc{W}_1 \times \mc{W}_1$ where two integer message sets $\mathcal{W}_1=\{1, 2, \ldots, \Delta_1\}$ and $\mathcal{W}_2=\{1, 2, \ldots, \Delta_2\}$ \cite{pradhan-choi-ram07,choi-pradhan08}.
We assume that there is some kind of correlation\footnote{Note that the meaning of correlation in the messages is different from the commonly used concept of correlation in the source coding problem.} between two message sets. More specifically, if the messages $W_1$ and $W_2$ are independent, then all possible pairs $(W_1, W_2)$ in the set $\mathcal{W}_1 \times \mathcal{W}_2$ are equally likely. On the other hand, if they are correlated, only some pairs $(W_1,W_2) \in A$ such that $A \subset \mathcal{W}_1 \times \mathcal{W}_2$ are equally likely and the other pairs have zero probability.

This correlation structure of the messages can be rephrased in terms of a bipartite graph $G$. The message pairs $(W_1,W_2) \in E(G)$ are equally likely with probability $\frac{1}{|E(G)|}$, and the message pairs $(W_1,W_2) \notin E(G)$ have zero probability  where a set $E(G) \subset \mc{W}_1 \times \mc{W}_2$, and each individual message $W_1$ and $W_2$ are individually
equally likely with probability $\frac{1}{|\mc{W}_1|}$ and $\frac{1}{|\mc{W}_2|}$, respectively. If the messages are independent, $E(G)=\mathcal{W}_1 \times \mathcal{W}_2$.

Therefore, a pair of correlated messages $(W_1,W_2)$ is an ordered tuple
$(\mc{W}_1,\mc{W}_2,G)$, which is characterized by two integer message sets $\mathcal{W}_1=\{1, 2, \ldots, \Delta_1\}$ and $\mathcal{W}_2=\{1, 2, \ldots, \Delta_2\}$, and an associated bipartite graph $G(\mc{W}_1,\mc{W}_2,E(G))$.

\begin{figure}[h]
\centering \epsfig{file=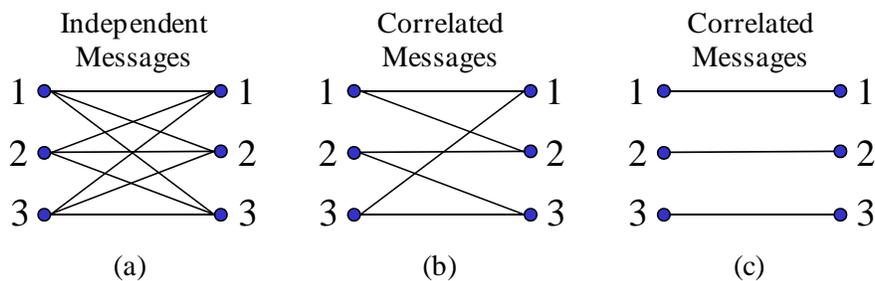, clip,
width=.7\linewidth} \centering \caption{\small Independent and correlated messages}\label{fig:correlation-ex}
\end{figure}

Let us consider a simple example shown in \cite{pradhan-choi-ram07}, which is illustrated in Fig. \ref{fig:correlation-ex}. Here, $\mathcal{W}_1=\mathcal{W}_2=\{1, 2, 3\}$. The vertices in the bipartite graph denote messages, and edges connecting two vertices imply that the corresponding message pair are equally likely. The complete bipartite graph of (a)
corresponds to the independent messages where all the possible pairs have equal probability $\frac{1}{9}$. However, (b) shows the correlated messages where each message pair in
$\{(1,1), (1, 2), (2, 2), (2, 3), (3, 3), (3, 1)\}$ has probability $\frac{1}{6}$, but (1,
3), (2, 1) and (3, 2) have zero probability. Moreover, (c) shows perfectly
correlated messages where only three message pairs (1, 1), (2, 2) and (3, 3)
can occur with the same probability $\frac{1}{3}$. The messages of (c) have higher correlation than those of (b).

\section{Problem Formulation and Summary of Result}\label{sec:summary}

In this section, we formulate the problem and show one of the main results of this paper. Consider a pair of correlated sources $X_1$ and $X_2$ with a joint probability distribution $p(x_1,x_2)$ and finite alphabets $\mc{X}_1$ and $\mc{X}_2$, respectively. In other words, a pair of correlated sources is an ordered tuple $(\mc{X}_1, \mc{X}_2, p(x_1,x_2))$. Let $(X_{11},X_{21})$, $(X_{12},X_{22}), \ldots$ be a sequence of jointly distributed random variables i.i.d. $\sim p(x_1,x_2)$. We denote $X_1^n \triangleq (X_{11},\ldots,X_{1n})$ and $X_2^n \triangleq  (X_{21},\ldots,X_{2n})$. We assume that the sources do not have a common part \cite{wyner75}.

\begin{figure}[h]\centering \epsfig{file=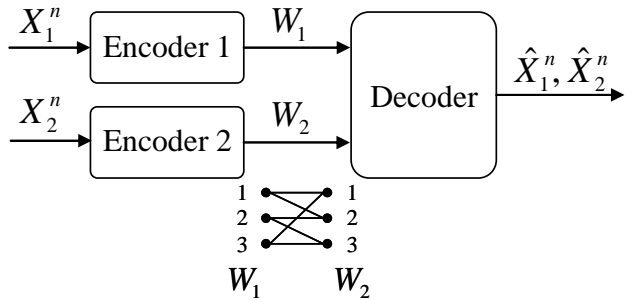, clip,
width=0.5\linewidth} \centering \caption{\small Distributed source coding with one distortion criterion and correlated messages} \label{fig:block-diagram}
\end{figure}

See the distributed source coding (DSC) system with correlated messages, shown in Fig. \ref{fig:block-diagram}, where the inputs of encoders are two discrete memoryless correlated sources $X_1$ and $X_2$.

The objective of this system is to represent the input $(X_1^n,X_2^n)$ into correlated messages $(W_1,W_2) \in \mc{W}_1 \times \mc{W}_2$ where $\mc{W}_i = \{1,2,\ldots,\Delta_i\}$ for $i=1,2$, and to reconstruct the original sources from the received messages $(W_1,W_2)$ under certain distortion conditions. Here, the correlated messages $(W_1,W_2)$ can be associated with nearly semi-regular graphs $G$ with parameters $(\Delta_1, \Delta_2, \Delta_1',\Delta_2', \mu)$. Let $\h{X}_1$ and $\h{X}_2$ be reconstructions of $X_1$ and $X_2$, respectively, where $X_1$ is lossless and $X_2$ is lossy with certain distortion criterion. Hence, $\mc{\h{X}}_1=\mc{X}_1$ and $\mc{\h{X}}_2$ may be different from $\mc{X}_2$.

\begin{dfn}
 The distortion measures between $(X_1^n,\h{X}_1^n)$ and $(X_2^n,\h{X}_2^n)$
  are defined by
  \begin{align}
    d_{x_1}(X_1^n,\h{X}_1^n) &= \frac{1}{n}\sum_{k=1}^n d_{x_1}(X_{1k},\h{X}_{1k}), \\
    d_{x_2}(X_2^n,\h{X}_2^n) &= \frac{1}{n}\sum_{k=1}^n d_{x_2}(X_{2k},\h{X}_{2k}),
  \end{align}
respectively, where the distortion measures $d_{x_1}$ and $d_{x_2}$ are any functions such that $d_{x_1}: \mc{X}_1 \times \mc{\h{X}}_1 \ra \{0,1\}$ and $d_{x_2}: \mc{X}_2 \times \mc{\h{X}}_2 \ra \mb{R}^{+}$. Let $d_{x_1}(x_1,\h{x}_1)=1-\delta_{x_1 \h{x}_1}$ where $\delta_{x_1 \h{x}_1}=1$ if $x_1=\h{x}_1$ and $\delta_{x_1 \h{x}_1}=0$ if $x_1 \neq \h{x}_1$.
\end{dfn}

A similar multiterminal source coding problem with one distortion criterion was studied by Berger and Yeung \cite{berger-yeung89}. So, this problem is also called the \emph{Berger-Yeung} problem. However, note that they only considered \emph{independent} messages as outputs of encoders. In our problem, we consider \emph{correlated} messages where the correlation structure is captured by a bipartite graph as illustrated in Section \ref{sec:prelim}. Hence, we also refer to our problem as ``Berger-Yeung problem with correlated messages''.

Now, let us define a DSC system with one distortion criterion and correlated messages as follows.
\begin{dfn}\label{def:dsc-system}
An $(n,\tau_{x_1},\tau_{x_2})$-DSC system for a pair of
correlated sources $(X_1,X_2)$ and a nearly semi-regular bipartite graph $G(\Delta_1, \Delta_2, \Delta_1',\Delta_2', \mu)$ is an ordered tuple $(f_1,f_2,g)$,
consisting of two encoding mappings $f_1$ and $f_2$, and one decoding mapping
$g$:
\begin{itemize}
  \item $f_1: \mc{X}_1^n \ra V_1(G)$, $f_2: \mc{X}_2^n \ra V_2(G)$,
  \item $g: E(G) \ra \mc{X}_1^n \times \mc{\h{X}}_2^n$,
  \item such that a performance measure is given by the
average distortions $\tau_{x_1}=E_{x_1} d_{x_1}(X_1^n,\h{X}_1^n)$ and $\tau_{x_2}=E_{x_2} d_{x_2}(X_2^n,\h{X}_2^n)$ where
 $(\h{X}_1^n,\h{X}_2^n)=g(f_1(X_1^n),f_2(X_2^n))$.
\end{itemize}
\end{dfn}

We also define achievable rate-distortion tuples for this problem as follows.
\begin{dfn}
  A rate-distortion tuple $(R_1$, $R_2$, $R_1'$, $R_2'$, $D)$ is said to be \emph{achievable} for a pair of correlated sources $(\mc{X}_1,\mc{X}_2,p(x_1,x_2))$,
 if for any $\e > 0$, and for all sufficiently large $n$,
  there exists a bipartite graph $G(\Delta_1, \Delta_2, \Delta_1', \Delta_2',\mu)$ and an
  associated $(n,\tau_{x_1},\tau_{x_2})$-DSC system as defined above such that:
  $\frac{1}{n}\log \Delta_i < R_i + \e$ for $i=1,2$,
  $\frac{1}{n}\log \Delta_i' < R_i' + \e$ for $i=1, 2$,
  $\frac{1}{n}\log \mu < \e$
  and the corresponding average distortions $\tau_{x_1} \leq \e$ and $\tau_{x_2} \leq D +\e$.
\end{dfn}
The goal is to find the achievable rate-distortion region $\mc{R(\mc{D})}$ which
is the set of all achievable rate-distortion tuples $(R_1, R_2, R_1',
R_2', D)$. We have found an information-theoretic characterization of $\mc{R(\mc{D})}$, which is one of the main results of this paper.
\vspace{5pt}

\begin{thm}\label{thm:main}
$\mc{R(\mc{D})}^* = \mc{R(\mc{D})}$ where
\begin{align}
\mc{R(\mc{D})}^* =\bigcup_{p(v|x_1,x_2)}\{(R_1, R_2, &R_1', R_2',D): \notag\\
 R_1 &\geq R_i' \geq 0 ~~\mbox{for $i=1,2$}, \label{eq:sc-main1}\\
 R_1' &\geq H(X_1|V), \label{eq:sc-main2}\\
 R_2' &\geq I(V;X_2|X_1), \label{eq:sc-main3}\\
 R_1+R_2'&= R_1'+R_2 \geq H(X_1)+I(V;X_2|X_1) \label{eq:sc-main4} \}
\end{align}
where $V$ is (i) an auxiliary random variable with finite alphabet $\mc{V}$ satisfying $|\mc{V}| \leq |\mc{X}_2| + 2$, and $p(x_1,x_2,v)=p(x_1,x_2)p(v|x_2)$ forms the Markov chain $X_1 \ra X_2 \ra V$, and (ii) there exists
$\h{X}_2(X_1,V)$ such that $E_{x_2} d_{x_2} (X_2,\h{X}_2) \leq D$.
\end{thm}

\begin{rem}
  Note that if we choose $V=X_2$, then we get Theorem 3 in \cite{pradhan-choi-ram07}, that is the achievable rate region for lossless distributed source coding for a pair of correlated sources using graphs, i.e., $D_{x_1}=D_{x_2}=0$, given by
  \begin{align}
   R_1 &\geq R_i' \geq 0 ~~\mbox{for $i=1,2$},\\
   R_1' &\geq H(X_1|X_2), \\
   R_2' &\geq H(X_2|X_1), \\
   R_1+R_2'&= R_1'+R_2 \geq H(X_1,X_2).
  \end{align}
\end{rem}

\begin{rem}
  Note also that there is a close connection between lossy distributed source coding problem using graphs \cite{choi09} and this problem. In \cite {choi09} where both sources are lossy, an inner bound to the achievable rate-distortion region is given by:
  \begin{align}
\bigcup_{p(u,v|x_1,x_2)}\{(R_1, R_2, &R_1', R_2',D_{x_1},D_{x_2}): \notag\\
 R_1 &\geq R_i' \geq 0 ~~\mbox{for $i=1,2$}, \\
 R_1' &\geq I(U;X_1|V), \\
 R_2' &\geq I(V;X_2|U), \\
 R_1+R_2'&= R_1'+R_2 \geq I(X_1,X_2;U,V) \}
\end{align}
where $U$ and $V$ are (1) auxiliary random variables with finite alphabets $\mc{U}$ and $\mc{V}$,  respectively, and $p(x_1,x_2,u,v)=p(x_1,x_2)p(u|x_1)p(v|x_2)$ forms the Markov chain $U \ra X_1 \ra X_2 \ra V$, and (2) there exist $\h{X}_1(U,V)$ and $\h{X}_2(U,V)$ such that $E_{x_i} d_{x_i} (X_i,\h{X}_i) \leq D_{x_i}$ for $i=1,2$.
If we choose $U=X_1$ in the Theorem 1 in \cite{choi09}, which is shown above, then we can obtain Theorem \ref{thm:main}.
\end{rem}

\begin{rem}
Theorem \ref{thm:main} gives only a partial characterization of the set of all nearly semi-regular bipartite graphs which can represent the given pair of correlated sources with certain amount of distortion.
\end{rem}

\section{Proof of Theorem \ref{thm:main}}\label{sec:thm-proof}

In this section, we present the proof of Theorem \ref{thm:main}. The proof consists of two parts: (1) the achievability of $\mc{R(\mc{D})}^*$ showing $\mc{R(\mc{D})}^* \subset \mc{R(\mc{D})}$ and (2) the converse part showing $\mc{R(\mc{D})}^* \supset \mc{R(\mc{D})}$.

\subsection{Achievability of $\mc{R(\mc{D})}^*$}\label{sec:ach-proof}
The proof of this achievability is similar to that of \cite[Theorem 1]{choi09}. We use the random binning technique used by Berger \cite{berger}, the concept of super-bin \cite{choi-pradhan08}, and the notion of strongly jointly typical sequences.

Given a pair of correlated sources $(X_1,X_2)$ with distribution $p(x_1,x_2)$, consider an auxiliary random variable $V$ which satisfies the conditions (i) and (ii) in Theorem \ref{thm:main}. Let us consider a fixed distribution $p(x_1,x_2,v)=p(x_1,x_2)p(v|x_2)$. Also, fix $\e >0$, and an integer $n\geq 1$. Let us choose $(R_1,R_2,R_1',R_2')$ as follows. $R_1 \geq H(X_1)-I(X_1;V)+\e$, $R_2 \geq I(V;X_2)-I(X_1;V)+\e$, and $R_1+R_2'=R_1'+R_2=A$ where $A=H(X_1)+I(V;X_2)-I(X_1;V)+2\e$.

\begin{figure}[h]
\centering \epsfig{file=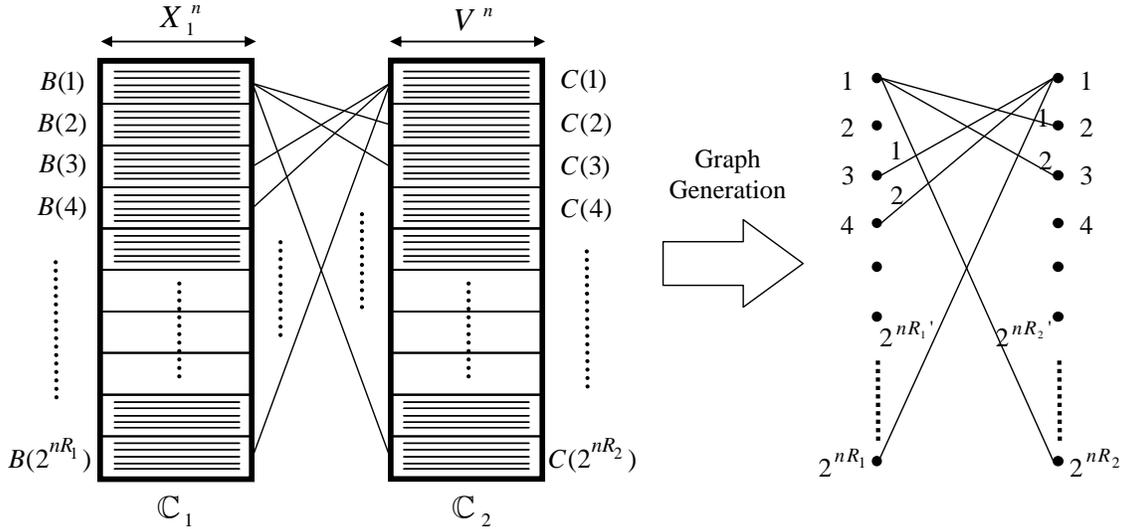, clip=,
width=0.9\linewidth} \centering \caption{\small A bipartite graph $\mb{G}(2^{nR_1}$, $2^{nR_2}$,
$2^{nR_1'}$, $2^{nR_2'}$, $\mu)$ generated from random codebooks $\mb{C}_1$ and $\mb{C}_2$.}
\label{fig:bin-index-graph}
\end{figure}

\noindent \textbf{Codebook Generation:}
First, draw $2^{n(H(X_1)+\e)}$ $n$-length sequences $X_1^n(k)$, for $k \in \{1,2,\ldots,
2^{n(H(X_1)+\e)}\}$, independently from $\Ae(X_1)$ with probability $\frac{1}{|\Ae(X_1)|}$ where $\Ae(X_1)$ is the strongly $\e$-typical set with respect to the distribution $p(x_1)$, which is the marginal of $p(x_1,x_2,v)$. Call this collection $\mb{C}_1$. Similarly, generate $2^{n(I(V;X_2)+\e)}$ sequences $V^n(l)$, for $l \in \{1,2,\ldots,2^{n(I(V;X_2)+\e)}\}$, from $\Ae(V)$, and call this collection $\mb{C}_2$. Then, divide $\mb{C}_1$ into $2^{nR_1}$ equal-size bins $B(i)$ for $i \in \{1, 2,\ldots, 2^{nR_1}\}$. Similarly, generate $C(j)$ for $j \in \{1,2\ldots, 2^{nR_2}\}$ from $\mb{C}_2$. This step is illustrated in Fig. \ref{fig:bin-index-graph}, where solid lines in bins $B(i)$ and $C(j)$ denote $n$-length sequences $X_1^n$ and $V^n$, respectively.

\noindent \textbf{Graph Generation:}
As shown in Fig. \ref{fig:bin-index-graph}, a random bipartite graph $\mb{G}$ can be generated from the bin indices of codebooks $\mb{C}_1$ and $\mb{C}_2$ and jointly typicality as follows. (1) $V_1(\mb{G})=\{1,2, \ldots, 2^{nR_1}\}$ and  $V_2(\mb{G})=\{1,2, \ldots, 2^{nR_2}\}$,
(2) $\forall (i,j) \in V_1(\mb{G}) \times V_2(\mb{G})$, $(i,j) \in E(\mb{G})$ if and only if there exists  at least one $\e$-strongly jointly
typical sequence pair $(X_1^n,V^n)$ in $B(i) \times C(j)$.

\emph{Encoding Error Events Due to the Degree Condition:}
Before the encoding and decoding steps, let us make sure that the generated graph $\mb{G}$ satisfies certain requirements. If the vertices of $\mb{G}$ do not satisfy this degree
requirements, the sources may not be able to be reliably represented by using this graph. So, an encoding error will be declared if either one of the following events occur:
\begin{itemize}
  \item $E_1$: $\exists i \in V_1(\mb{G})$ such that $\left|
  \frac{1}{n}\log{\rm deg}_{\mb{G},1}(i)-R_2' \right| > \e'$,
  \item $E_2$: $\exists j \in V_2(\mb{G})$ such that $\left|
  \frac{1}{n}\log{\rm deg}_{\mb{G},2}(j)-R_1' \right| >   \e'$,
\end{itemize}
where $\e_1(\e)$ is a continuous function of $\e$ with
$\e_1(\e) \rightarrow 0$ as $\e \rightarrow 0$, and is characterized
in Appendix\ref{app:e1}, and $\e'> 3\e_1$.

\noindent \textbf{Choosing Message Correlation:}
If none of the above error events $E_1$ and $E_2$ occurs, then choose
$G=\Bbb{G}$. If any of the above two error events occurs, then pick
any nearly semi-regular graph with parameters $(2^{nR_1}$,$2^{nR_2}$,$2^{nR'_1}$,$2^{nR'_2}$,$2^{n\e'})$
and call it $G$ and no guarantee will be given regarding the
probability of decoding error.

For this graph $G$, and the given correlated sources $X_1$ and $X_2$, using
the above random codebooks $\Bbb{C}_1$ and $\Bbb{C}_2$, we construct
an $(n,\tau_{x_1},\tau_{x_2})$-DSC system, where $\tau_{x_1} \leq \e$ and $\tau_{x_2} \leq D +\e$.

\noindent \textbf{Encoding:}
If $(E_1\cup E_2)$ does not occur, for the given source $X_1^n$, encoder 1 looks for a sequence $x_1^n(k) \in \Bbb{C}_1$ such that $x_1^n(k)=X_1^n$, and sends the bin index $i$ such that $x_1^n(k) \in B(i)$. If there is no such index $i$, it sends any random index $i$ chosen uniformly from $\{1,2, \ldots, 2^{nR_1}\}$.

For the given source $X_2^n$, encoder 2 looks for a sequence $v^n(l)\in \Bbb{C}_2$ such that
$(X_2^n,v^n(l)) \in \Ae(X_2,V)$ where $\Ae(X_2,V)$ is the jointly strongly $\e$-typical set with respect to the distribution $p(x_2,v)$, and sends $j$ satisfying $v^n(l) \in C(j)$.
Let us denote the selected sequences $v^n(l)$ by $v^n(X_2^n)$. Then, from the Markov lemma \cite{berger,cover-thomas}, $(X_1^n,X_2^n,v^n(X_2^n))$ becomes jointly typical, i.e., $(X_1^n,X_2^n,v^n(X_2^n)) \in \Aetl(X_1,X_2,V)$ where $\tl{\e}=K\e$ for an appropriate constant $K$. If there is no such index $j$, it sends any random index $j$ chosen uniformly from $\{1,2, \ldots, 2^{nR_2}\}$.

\noindent \textbf{Decoding:} Given the received index pair $(i,j)$,
the decoder looks for the unique pair of sequences $(X_1^n,v^n(X_2^n))\in B(i) \times C(j)$ such that $(X_1^n,v^n(X_2^n)) \in \Aetl(X_1,V)$.
Then, it calculates $\h{X}_1^n$ and $\h{X}_2^n$ from $\h{X}_{1m}=X_{1m}$ and $\h{X}_{2m}=\h{X}_2(X_{1m},v_m(X_2^n))$ for $1 \leq m \leq n$.

If there exists any other $(x_1^n,v^n)\in B(i) \times C(j)$ such that $(x_1^n,v^n)\neq (X_1^n,v^n(X_2^n))$ and $(x_1^n,v^n) \in \Aetl(X_1,V)$, then an error is declared, and it sets $\h{X}_1^n=\h{x}_1^n$ and $\h{X}_2^n=\h{x}_2^n$ where $\h{x}_1^n$ and $\h{x}_2^n$ are arbitrary sequences in $\mc{X}_1^n$ and $\mc{\h{X}}_2^n$, respectively.

\noindent \textbf{Probability of Error Analysis:} Let $E$ denote the
error event. Then, the probability of error $P(E)$ can be given by
\begin{align}
  P(E)&=P(E_1 \!\cup\! E_2)P(E|E_1 \! \cup \! E_2)\!+\!P(E \cap E_1^c \cap E_2^c)\\
       &\leq P(E_1 \!\cup\! E_2)+P(E \cap E_1^c \cap E_2^c).\label{eq:error-prob}
\end{align}

By using the similar techniques shown in \cite[p.\ 2847]{choi-pradhan08}, it is can be shown that, for sufficiently large $n$, $P(E_1) < \frac{\e}{7}$, and  $P(E_2) < \frac{\e}{7}$,
if $R_2'=A-R_1$ and $R_1'=A-R_2$, respectively. This means that with high probability
we can obtain a nearly semi-regular bipartite graph $\mb{G}$ such that each vertex
in $V_1(\mb{G})$ has degree nearly equal to $2^{nR'_2}$
and each vertex in $V_2(\mb{G})$ has degree nearly equal to
$2^{nR'_1}$.

The second probability in (\ref{eq:error-prob}) can be bounded as given
in the following lemma.
\begin{lem} \label{lem:bc-cc-3}
For any $\epsilon>0$, and sufficiently large $n$, \beq P(E \cap
E_1^c \cap E_2^c) < \frac{5 \epsilon}{7} \eeq
\end{lem}

{\it Proof}: Now let us calculate the probability $P(E \cap E_1^c
\cap E_2^c)$. If previous error events $E_1$ or $E_2$ do not
occur, we define other error events as follows:
\begin{itemize}
  \item[$E_3$]: $(X_1^n,X_2^n) \notin \Ae(X_1,X_2)$,
  \item[$E_4$]: $\nexists X_1^n \in \mb{C}_1$,
  \item[$E_5$]: $\nexists v^n(X_2^n) \in \mb{C}_2$,
  \item[$E_6$]: $(X_1^n,X_2^n,v^n(X_2^n)) \notin \Aetl(X_1,X_2,V)$,
  \item[$E_7$]: $\exists (x_1^n,v^n) \neq (X_1^n,v^n(X_2^n))$ such that $(x_1^n,v^n) \in B(i) \times C(j)$ and $(x_1^n,X_2^n,v^n) \in \Aetl(X_1,X_2,V)$.
\end{itemize}
Then,
\begin{align}
  P(E \cap E_1^c \cap E_2^c)&=P\left( \cup_{i=3}^7 E_i \cap E_1^c \cap E_2^c \right)\\
                                   &\leq  \sum_{i=3}^7 P\left(E_i \cap E_1^c \cap E_2^c \right)
\end{align}

By the property of jointly typical sequences \cite{cover-thomas},
$P(E_3 \cap E_1^c \cap E_2^c)< \frac{\e}{7}$ for sufficiently
large $n$. $P(E_4 \cap E_1^c \cap E_2^c)< \frac{\e}{7}$ since $|\mb{C}_1| > 2^{nH(X_1)}$ and from the property of typical set \cite[p.\ 371]{berger}. $P(E_5 \cap E_1^c \cap E_2^c)< \frac{\e}{7}$ since $|\mb{C}_2| > 2^{nI(V;X_2)}$ \cite[Lemma 2.1.3]{berger}. $P(E_6 \cap E_1^c \cap E_2^c)< \frac{\e}{7}$ from the Markov lemma as described in \cite{berger,cover-thomas}. $P(E_7 \cap E_1^c \cap E_2^c)< \frac{\e}{7}$ since $R_1+R_2 \geq H(X_1)+I(V;X_2)-I(X_1;V)+2\e$ \cite{berger}.

Therefore, by applying the union bound we have
$P(E) \leq  P(E_1)+P(E_2)+P(E \cap E_1^c \cap E_2^c) \leq \e$.

\noindent \textbf{Calculation of Distortion:}
Now let us calculate the resulting distortion $d_{x_2}(X_2^n,\h{X}_2^n)$.
Following \cite{berger}, if the error event $E$ does not occur, $d_{x_2}(X_2^n,\h{X}_2^n) \leq D + \e^*$ from the jointly typicality of $(X_1^n,X_2^n,v^n(X_2^n))$ where $\e^* \ra 0$ as $\e \ra 0$.
So,
\begin{equation}
  d_{x_2}(X_2^n,\h{X}_2^n) \leq (1-P(E))(D + \e^*)+P(E)d_{\rm max}
\end{equation}
where $d_{\rm max}$ is the maximum distortion for any individual sequence.

Hence, for sufficiently large $n$, the distortions for $X_1^n$ and $X_2^n$ can be close to $0$ and $D$, respectively, if $P(E)$ is small.

In every realization of random codebooks, we have obtained a graph $G$ with parameters
 $(2^{nR_1}$,$2^{nR_2}$,$2^{nR'_1}$,$2^{nR'_2}$,$2^{n\e'})$, and averaged over the ensemble of random codebooks, the average distortions for $X_1^n$ and $X_2^n$ are close to $0$ and $D$, respectively.
 Therefore, the proof of the achievability is complete.
 \hfill $\blacksquare$

\subsection{The Converse}\label{sec:converse}
Now we prove the converse part of Theorem \ref{thm:main}. Some steps of the proof is similar to the converse proof in \cite{berger-yeung89}.

Let us assume a rate-distortion tuple $(R_1$, $R_2$, $R_1'$, $R_2'$, $D)$ is achievable. Then  for any $\e > 0$, and for all sufficiently large $n$,
  there exists a bipartite graph $G(\Delta_1, \Delta_2, \Delta_1', \Delta_2',\mu)$ and an
  associated $(n,\tau_{x_1},\tau_{x_2})$-DSC system as defined in Definition \ref{def:dsc-system} such that:
  $\frac{1}{n}\log \Delta_i < R_i + \e$ for $i=1,2$,
  $\frac{1}{n}\log \Delta_i' < R_i' + \e$ for $i=1, 2$,
  $\frac{1}{n}\log \mu < \e$
  and the corresponding average distortions $\tau_{x_1} \leq \e$ and $\tau_{x_2} \leq D +\e$. Let $f_1(X_1^n)=W_1$ and  $f_2(X_2^n)=W_2$.

 Then, from the vector version of Fano's inequality \cite[Lemma 1]{berger-yeung89}, we can have
  \begin{align}
    H(X_1^n|\h{X}_1^n) \leq n\tau_{x_1} \log(|\mc{X}_1|-1)+nH_b(\tau_{x_1}) \leq n\e_n \label{eq:1}
  \end{align}
  where $\e_n \ra 0$ as $n \ra \infty$ and $H_b(\cdot)$ is the binary entropy function.

Using the constraints on the degree of the vertices in the nearly semi-regular graph associated with correlated messages, we have
\begin{align}
  nR_1' &\geq H(W_1|W_2) - \log \mu \\
        &\geq I(W_1;X_1^n|W_2) - \log \mu \\
        &= H(X_1^n|W_2)-H(X_1^n|W_1,W_2) - \log \mu \\
        &\overset{(a)}{=} H(X_1^n|W_2)-H(X_1^n|W_1,W_2,\h{X}_1^n) - \log \mu \\
        &\overset{(b)}{\geq} \sum_{k=1}^n H(X_{1k}|W_2,X_1^{k-1})-H(X_1^n|\h{X}_1^n) - \log \mu \\
        &\geq \sum_{k=1}^n H(X_{1k}|W_2,X_1^{k-1},X_{1(k+1)},\ldots,X_{1n})-H(X_1^n|\h{X}_1^n) - \log \mu \\
        &\overset{(c)}{\geq} \sum_{k=1}^n H(X_{1k}|V_k) - n\e_n - \log \mu
\end{align}
where (a) is from $I(X_1^n;\h{X}_1^n|W_1,W_2)=0$ due to the Markov chain $X_1^n \ra (W_1,W_2) \ra \h{X}_1^n$, (b) is obtained by using the chain rule and removing conditioning, and (c) follows from Fano's inequality (\ref{eq:1}) and by defining $V_k=(W_2,X_1^{k-1},X_{1(k+1)},\ldots,X_{1n})$.

Thus, we have
\begin{align}
  R_1' \geq \frac{1}{n}\sum_{k=1}^n H(X_{1k}|V_k) - \e_n - \frac{1}{n}\log \mu. \label{eq:4}
\end{align}

Similarly, we also have
\begin{align}
  nR_2' &\geq H(W_2|W_1) - \log \mu \label{eq:3}\\
        &= H(W_2)-I(W_1;W_2) - \log \mu \\
        &\overset{(a)}{\geq} H(W_2)-I(X_1^n;W_2) - \log \mu \\
        &= H(W_2|X_1^n) - \log \mu \\
        &\geq I(X_2^n;W_2|X_1^n) - \log \mu \\
        &= H(X_2^n|X_1^n)-H(X_2^n|X_1^n,W_2) -\log \mu \\
        &\overset{(b)}{=} \sum_{k=1}^n [H(X_{2k}|X_{1k})-H(X_{2k}|W_2,X_1^n,X_2^{k-1})] - \log \mu \\
        &\overset{(c)}{\geq} \sum_{k=1}^n [H(X_{2k}|X_{1k})-H(X_{2k}|W_2,X_1^n)] - \log \mu \\
        &\overset{(d)}{=} \sum_{k=1}^n [H(X_{2k}|X_{1k})- H(X_{2k}|V_k,X_{1k})] - \log \mu \\
        &= \sum_{k=1}^n I(V_k;X_{2k}|X_{1k}) - \log \mu \label{eq:2}
\end{align}
where (a) is from $I(W_1;W_2) \leq I(X_1^n;W_2)$ due to the Markov chain $W_2 \ra X_1^n \ra W_1$, (b) is obtained by using the chain rule and memoryless property of the sources $(X_1,X_2)$, (c) is from removing conditioning, and (d) is obtained by defining $V_k=(W_2,X_1^{k-1},X_{1(k+1)},\ldots,X_{1n})$.

So, we have
\begin{align}
  R_2' \geq \frac{1}{n}\sum_{k=1}^n  I(V_k;X_{2k}|X_{1k}) -  \frac{1}{n}\log \mu. \label{eq:5}
\end{align}

For the sum-rate $R_1+R_2'=R_1'+R_2$, using the constraints on the size of the edge set and on the degree of the vertices of the nearly semi-regular graph associated with correlated messages, i.e., $(W_1,W_2) \in E(G)$ and $|E(G)|\leq \mu 2^{n(R_1+R_2')}$, we have
\begin{align}
  &n(R_1+R_2')=n(R_1'+R_2) \notag \\
   &\geq H(W_1,W_2) - \log \mu \\
        &= H(W_1)+H(W_2|W_1) - \log \mu \\
        &= I(X_1^n;W_1)+\underbrace{H(W_1|X_1^n)}_{=0} + H(W_2|W_1) -\log \mu \\
        &\overset{(a)}{=} H(X_1^n)-H(X_1^n|W_1) + H(W_2|W_1)- \log \mu \\
        &\overset{(b)}{\geq} H(X_1^n)-n\e_n + H(W_2|W_1)- \log \mu \\
        &\overset{(c)}{=} \sum_{k=1}^n [H(X_{1k})+ I(V_k;X_{2k}|X_{1k})]- n\e_n - \log \mu
\end{align}
where (a) follows the fact that $H(W_1|X_1^n)=0$ since $f_1(X_1^n)=W_1$, (b) is obtained from $H(X_1^n|W_1) \leq n\e_n$ due to the Fano's inequality \cite{cover-thomas},  (c) follows from the chain rule and the memoryless property of the source $X_1$ and (\ref{eq:3}) and (\ref{eq:2}) above, i.e., $H(W_2|W_1) \geq \sum_{k=1}^n I(V_k;X_{2k}|X_{1k})$.

Hence, we have
\begin{align}
  &R_1+R_2'=R_1'+R_2 \\
          &\geq \frac{1}{n}\sum_{k=1}^n [H(X_{1k})+ I(V_k;X_{2k}|X_{1k})] - \e_n - \frac{1}{n}\log \mu. \label{eq:6}
\end{align}
Therefore, we can have the converse by taking the limit of inequalities (\ref{eq:4}), (\ref{eq:5}) and (\ref{eq:6}) as $n \ra \infty$. \hfill $\blacksquare$

\section{Representation of a Pair of Correlated Sources into Different Graphs}\label{sec:different}

\begin{figure}[!h] \centering
\epsfig{file=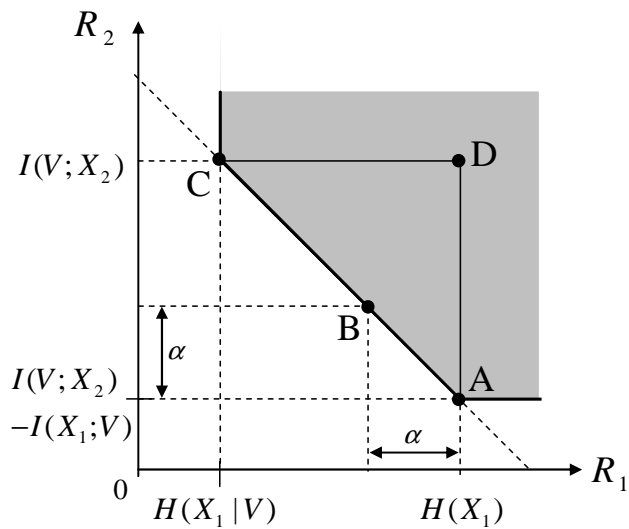, clip=, width=.5\linewidth}
\centering \caption{\small The achievable rate-distortion region of DSC with one distortion criterion in terms of $R_1$ and $R_2$}\label{fig:rate-region}
\end{figure}

In this section, we show that in the case of DSC with one distortion criterion, a pair correlated sources $(X_1,X_2)$ can be reliably represented into many different graphs without increasing the redundancy.  In other words, if $(R_1,R_2)$ are inside the triangle $ACD$ in Fig. \ref{fig:rate-region} and $R_1'$ and $R_2'$ satisfy the condition $R_1+R_2'=R_1'+R_2=H(X_1)+I(V;X_2|X_1)$, then many different graphs $G$ with parameters  $(2^{nR_1}$,$2^{nR_2}$,$2^{nR'_1}$,$2^{nR'_2}$,$2^{n\e'})$ can represent the same sources $(X_1,X_2)$. Note that this is similar to the lossless DSC case shown in \cite[Section \ 5]{pradhan-choi-ram07}.

Let us consider some special cases as follows.
\begin{itemize}
  \item Point $A$: $R_1=R_1'=H(X_1)$, $R_2=R_2'=I(V;X_2|X_1)=I(V;X_2)-I(X_1;V)$. In this case, we get a nearly complete graph, which is an efficient representation of the sources that has the least redundancy in the conventional sense, i.e., in terms of $(R_1,R_2)$. For this point, the bin sizes $|B(i)|$ and $|C(j)|$ in Fig. \ref{fig:bin-index-graph} are roughly unity and $2^{nI(X_1;V)}$, respectively.
  \item Point $B$: $R_1=R_1'=H(X_1)-\alpha$, $R_2=R_2'=I(V;X_2)-I(X_1;V)+\alpha$ for $0 < \alpha < I(X_1;V)$. This case corresponds to an arbitrary point $(R_1,R_2)$ on the line segment $AC$. This also gives a nearly complete graph, where $|B(i)|$ and $|C(j)|$ in Fig. \ref{fig:bin-index-graph} are roughly $2^{n\alpha}$ and $2^{n(I(X_1;V)-\alpha)}$, respectively.
  \item Point $C$: $R_1=R_1'=H(X_1|V)=H(X_1)-I(X_1;V)$, $R_2=R_2'=I(V;X_2)$. This also gives a nearly complete graph, where $|B(i)|$ and $|C(j)|$ in Fig. \ref{fig:bin-index-graph} are roughly $2^{nI(X_1;V)}$ and unity, respectively.
  \item Point $D$: $R_1=H(X_1)$, $R_2=I(V;X_2)$, $R_1'=H(X_1|V)=H(X_1)-I(X_1;V)$, $R_2'=I(V;X_2|X_1)=I(V;X_2)-I(X_1;V)$.  In this case, the graph has the maximum redundancy in the conventional sense, and is not complete. However, this is also an efficient representation since the total number of edges of the graph is nearly equal to $2^{n(H(X_1)+I(V;X_2)-I(X_1;V))}$, where $|B(i)|$ and $|C(j)|$ in Fig. \ref{fig:bin-index-graph} are roughly unity and unity, respectively.
\end{itemize}

Therefore, an arbitrary point $(R_1,R_2)$ on the line segment $AC$ (such as point $A$, $B$ and $C$) has the least redundancy and point $D$ has the maximum redundancy in the conventional sense. For every point in the triangle $ACD$ in Fig.\ref{fig:rate-region}, we can obtain an equally efficient representation of the correlated sources $(X_1,X_2)$ into a nearly semi-regular graph. Here, equally efficient representation means that the cardinalities
of edge sets of these graphs for the different values of $(R_1,R_2)$ are nearly the same.

\section{Functional Duality Between Berger-Yeung Problem with Correlated Messages and Semi-deterministic Broadcast Channel with Correlated Messages}\label{sec:duality}
In this section, we discuss functional duality between Berger-Yeung problem (BYP) with correlated messages and semideterministic broadcast channel (SBC) with correlated messages. We show that, under certain conditions, for a given BYP with correlated messages problem, a dual SBC with correlated messages problem can be obtained where both problems have the same joint distribution and the same correlation structure in the messages, and vice versa.
Before discussing the functional duality, we briefly recall BYP with correlated messages in Section \ref{sec:summary} and SBC with correlated messages in \cite[p.\ 2848]{choi-pradhan08}. In our discussion, we denote the given distribution and the distribution which optimizes a given objective function by $\bar{p}(\cdot)$ and $p^*(\cdot)$, respectively.

\subsection{Berger-Yeung Problem with Correlated Messages}

\begin{figure}[h]
\centering \epsfig{file=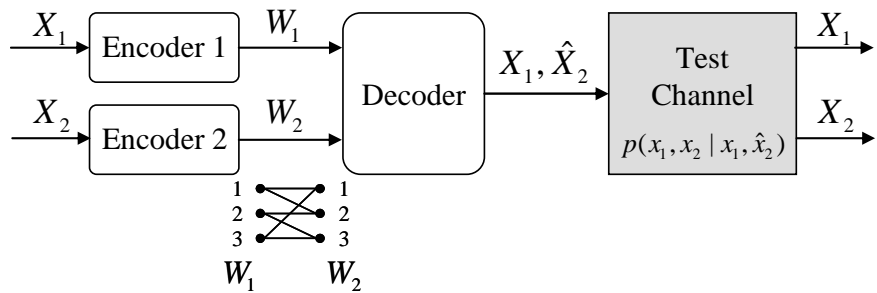, clip,
width=.7\linewidth} \centering \caption{\small Berger-Yeung problem with correlated messages} \label{fig:dsc-block}
\end{figure}

Consider a BYP with correlated messages, shown in Fig. \ref{fig:dsc-block}, where $X_1$ and $X_2$ are two correlated discrete memoryless stationary sources  with a given joint probability distribution $\bar{p}(x_1,x_2)$, and with finite alphabets $\mc{X}_1$ and $\mc{X}_2$, respectively. The objective of this system is to represent $(X_1^n,X_2^n)$ into correlated messages $(W_1,W_2) \in \mc{W}_1 \times \mc{W}_2$ such that $\mc{W}_i = \{1,2,\ldots,\Delta_i\}$ for $i=1,2$, which can be associated with nearly semi-regular graphs $G(\Delta_1, \Delta_2, \Delta_1',\Delta_2', \mu)$ and to reconstruct the original sources, $(X_1^n,\h{X}_2^n)$, from the graphs under certain distortion conditions. Here, the encoders do not communicate with each other.

Let $d: \mc{X}_1 \times \mc{X}_2 \times \mc{\h{X}}_2 \ra \mb{R}^{+}$ be the joint distortion measure, where $\mc{\h{X}}_2$ is the reconstruction alphabet of $X_2$.
The encoders are given by $\mc{X}_i^n \ra V_i(G)$ for $i=1,2$, and the decoder is given by $E(G) \ra \mc{X}_1^n \times \mc{\h{X}}_2^n$. An achievable rate-distortion region $\mc{R}(D)_{BYP}$ for a distortion constraint $D$ is given by
\begin{align}
\mc{R}(D)_{BYP} =\bigcup_{p(v|x_1,x_2)}\{(&R_1, R_2, R_1', R_2',D): \notag\\
 &R_1 \geq R_i' \geq 0 ~~\mbox{for $i=1,2$}, \\
 &R_1' \geq H(X_1|V), \\
 &R_2' \geq I(V;X_2|X_1), \\
 R_1+&R_2'= R_1'+R_2 \geq H(X_1) + I(V;X_2|X_1) \}
\end{align}
where $V$ is an auxiliary random variable with finite alphabets $\mc{V}$ satisfying $|\mc{V}| \leq |\mc{X}_2| + 2$, and $p(x_1,x_2,v)=\bar{p}(x_1,x_2)p(v|x_2)$ forms the Markov chain $X_1 \ra X_2 \ra V$, and there exists $\h{X}_2(X_1,V)$ such that $Ed \leq D$ and $X_2 \ra V \ra (X_1,\h{X}_2)$.

Note that $H(X_1|V)=H(X_1)-I(X_1;V)$ and $I(V;X_2|X_1)=I(V;X_2)-I(X_1;V)$ due to the Markov chain $X_1 \ra X_2 \ra V$. So, the sum-rate $R_{BYP}^{sum}(D)=R_1+R_2'= R_1'+R_2$ can be also expressed as
\begin{align}
  R_{BY}^{sum}(D) = \min [H(X_1)+I(V;X_2)-I(X_1;V)]
\end{align}
where the minimization is taken over $p(v|x_2)$ and $p(\h{y}_2|v,x_1)$.

\subsection{Semideterministic Broadcast Channel with Correlated Messages \cite{choi-pradhan08}}

\begin{figure}[h]
\centering \epsfig{file=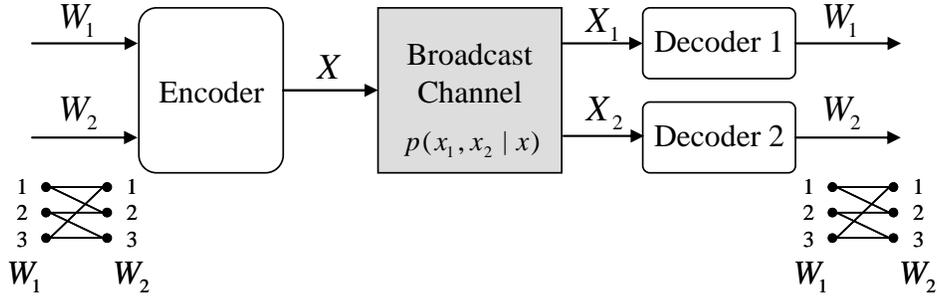, clip,
width=.75\linewidth} \centering \caption{\small Semideterministic broadcast channel with correlated messages} \label{fig:bcc-block}
\end{figure}

Consider a general discrete memoryless stationary SBC system with correlated messages, shown in Fig. \ref{fig:bcc-block}, with a given conditional distribution $\bar{p}(x_2|x)$ and a deterministic function $x_1=f(x)$ where $\mc{X}$ is input alphabet and $\mc{X}_i$ are output alphabets, respectively, for $i=1,2$.
The objective of this system is to send simultaneously a pair of correlated messages $(W_1,W_2)$, which can be associated with graphs $G(\Delta_1, \Delta_2, \Delta_1',\Delta_2', \mu)$, to the two receivers over the channel where $(W_1,W_2) \in \mc{W}_1 \times \mc{W}_2$ such that $\mc{W}_i = \{1,2,\ldots,\Delta_i\}$ for $i=1,2$. Here, the decoders do not communicate with each other. We assume that there is no common message in the two messages.

Let $w: \mc{X} \ra \mb{R}^{+}$ be the input cost measure associated with this channel.
The encoder is a mapping $E(G) \ra \mc{X}^n$, and the decoders are given by  $\mc{X}_i^n \ra V_i(G)$ for $i=1,2$.

The capacity region $\mc{R}(W)_{SBC}$ for a cost constraint $W$ is given by
\begin{align}
\mc{R}(W)_{SBC} =\bigcup_{p(v,x)}\{(R_1, &R_2, R_1', R_2'):\notag\\
 &R_1 \geq R_i' \geq 0 ~~\mbox{for $i=1,2$}, \label{eq:bc-main1}\\
 &R_1 \leq H(X_1), \\
 &R_2 \leq I(V;X_2), \\
 R_1\!+\!&R_2'\!=\! R_1'\!+\!R_2 \leq H(X_1|V)\!+\!I(V;X_2) \}
\end{align}
such that $Ew \leq W$, $V \ra X \ra (X_1,X_2)$ where $V$ is an auxiliary random variable with finite alphabet $\mc{V}$ satisfying $|\mc{V}| \leq |\mc{X}| + 2$, and $p(v,x)=p(v)p(x|v)$ forms the Markov chain $X_2 \ra V \ra X$.

The sum-rate $R_{SBC}^{sum}(W)=R_1+R_2'= R_1'+R_2$ is given by
\begin{align}
  R_{SBC}^{sum}(W) \!=\! \max [H(X_1) \!+\!I(V;X_2)\!-\!I(X_1;V)]
\end{align}
where the maximization is taken over $p(v,x)$.

\begin{figure}[h]
\centering \epsfig{file=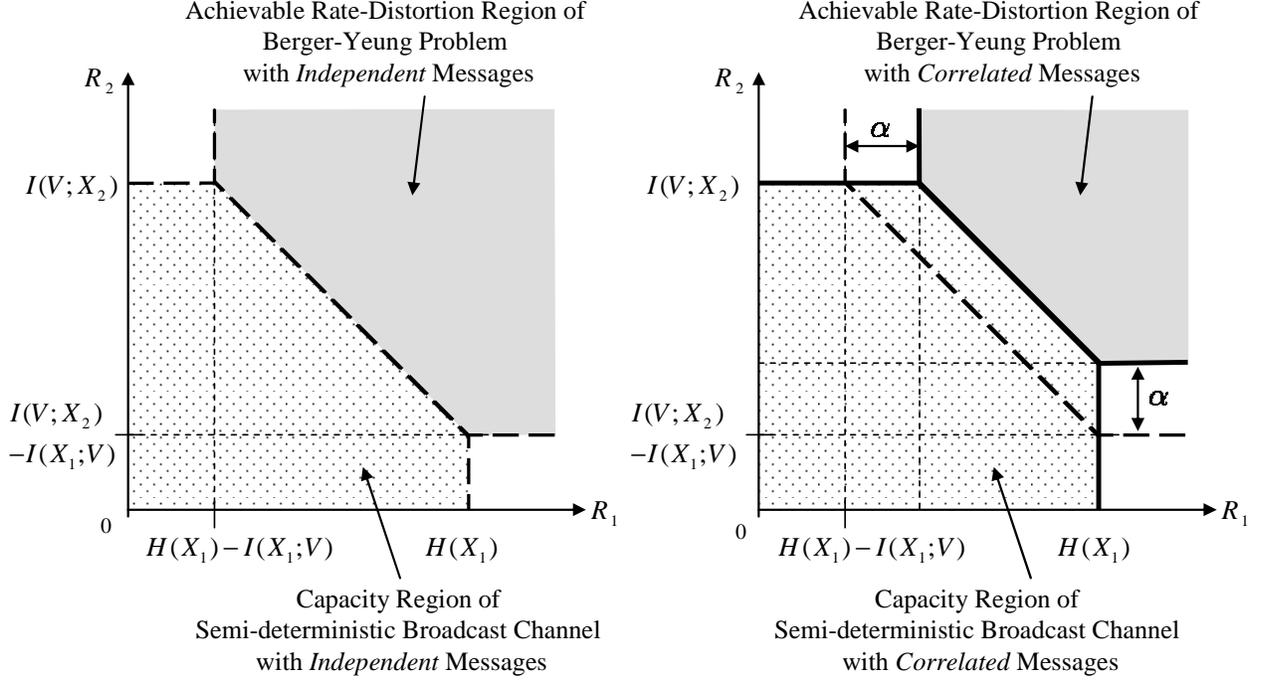, clip,
width=1.0\linewidth} \centering \caption{\small Achievable rate-distortion region of BYP and the capacity region of SBC with independent and correlated messages} \label{fig:two-rate-region}
\end{figure}

The achievable rate-distortion region of BYP and the capacity region of SBC with independent and with correlated messages are depicted in Fig. \ref{fig:two-rate-region}, where $R_i'=R_i-\alpha$ for $i=1,2$ such that $0 \leq \alpha \leq \min\{R_1,R_2\}$. Note that $\alpha=0$ for independent messages and $\alpha>0$ for correlated messages.

\subsection{Functional Duality between BYP and SBC with Correlated Messages}\label{subsec:duality}

Now, we discuss functional duality between BYP with correlated messages and SBC with correlated messages. We show that, under certain conditions, for a given BYP with correlated messages, a dual SBC with correlated messages can be obtained where both problems have the same joint distribution and the same correlation structure in the messages, and vice versa.

The following theorem is one of the main results of this paper.

\begin{thm}\label{thm:main1}
(1) For a given BYP with correlated messages $(W_1,W_2)$, which can be associated with a graph $G(\Delta_1, \Delta_2, \Delta_1',\Delta_2', \mu)$, a given source $(X_1,X_2) \sim \bar{p}(x_1,x_2)$ with alphabets $\mc{X}_i$ for $i=1,2$ and reconstruction alphabet $\mc{\h{X}}_2$, a distortion measure  $d: \mc{X}_1 \times \mc{X}_2 \times \mc{\h{X}}_2 \ra \mb{R}^{+}$, and a distortion constraint $D$, suppose $\{p^*(v|x_2),p^*(\h{x}_2|x_1,v)\}$ achieves the minimum of the sum-rate distortion function $R_{BYP}^{sum}(D)$:
\begin{align}
  R_{BYP}^{sum}(D) \!=\! \min [H(X_1)\!+\!I(V;X_2)\!-\!I(X_1;V)]
\end{align}

such that $Ed \leq D$, $X_1 \ra X_2 \ra V$ and $X_2 \ra V \ra (X_1,\h{X}_2)$. Then, $\bar{p}(x_1,x_2)$ and $\{p^*(v|x_2),p^*(\h{x}_2|x_1,v)\}$ induce the following joint distribution:
\begin{align}
p^*(x_1,x_2,v,\h{x}_2)=\bar{p}(x_1,x_2)p^*(v|x_2)p^*(\h{x}_2|x_1,v).
\end{align}
Let $p^*(v)$, $p^*(x_1,\h{x}_2|v)$ and $p^*(x_2|x_1,\h{x}_2)$ be the corresponding marginals. If $p^*(x_1,x_2,v,\h{x}_2)$ satisfies $V \ra X \ra (X_1,X_2)$ where $X=(X_1,\h{X}_2)$, then $\exists$ a dual SBC problem for the channel $p^*(x_2|x)=p^*(x_2|x_1,\h{x}_2)$ and $x_1=f(x)$ with correlated messages $(W_1,W_2)$, which can be associated with a graph $G(\Delta_1, \Delta_2, \Delta_1',\Delta_2', \mu)$,  input alphabet $\mc{X}$ and output alphabets $\mc{X}_i$, for $i=1,2$, a cost measure $w:\mc{X} \ra \mb{R}^{+}$, and a cost constraint $W$ such that:
\begin{itemize}
  \item $R_{BYP}^{sum}(D)=R_{SBC}^{sum}(W)$, i.e., $\min [H(X_1)\!+\!I(V;X_2)\!-\!I(X_1;V)]= \max [H(X_1)\!+\!I(V;X_2)\!-\!I(X_1;V)]$ where the minimum is taken over $p(v|x_2)$ and $p(\h{x}_2|x_1,v)$ with the given source distribution $\bar{p}(x_1,x_2)$ such that $Ed \leq D$, $X_1 \ra X_2 \ra V$ and $X_2 \ra V \ra (X_1,\h{X}_2)$; and the maximum is taken over $p(v)$ and $p(x|v)$ with the fixed channel conditional distribution $p^*(x_2|x)$ and $x_1=f(x)$ such that $Ew \leq W$, $V \ra X \ra (X_1,X_2)$, and $X_2 \ra V \ra X$,
  \item the distributions $p^*(v)$ and $p^*(x|v)=p^*(x_1,\h{x}_2|v)$ obtained from the BYP achieve the maximum in the dual SBC problem,
  \item the correlation structure of the messages $(W_1,W_2)$ of the dual SBC problem is the same as that of the given BYP, i.e, the graph $G(\Delta_1, \Delta_2, \Delta_1',\Delta_2', \mu)$ of the dual SBC is the same as that of the given BYP
\end{itemize}
provided the cost measure and the cost constraint are chosen such that
\begin{equation}
  w(x)=c_1 D(p^*(x_1,x_2|x)||\bar{p}(x_1,x_2)) + \theta
\end{equation}
and $W=E_{p^*(x)}[w(X)]$ where $D(\cdot||\cdot)$ is the relative entropy \cite{cover-thomas}, $c_1 >0$ and $\theta$ are arbitrary constants.

(2) For a given SBC with correlated messages $(W_1,W_2)$, which can be associated with a graph $G(\Delta_1, \Delta_2, \Delta_1',\Delta_2', \mu)$, and a given channel conditional distribution $\bar{p}(x_2|x)$ and $x_1=f(x)$, input alphabet $\mc{X}$ and output alphabets $\mc{X}_i$  for $i=1,2$, a cost measure  $w: \mc{X} \ra \mb{R}^{+}$, and a cost constraint $W$, suppose $\{p^*(v),p^*(x|v)\}$ achieves the maximum of the sum-rate cost function $R_{SBC}^{sum}(W)$:
\begin{equation}
  R_{SBC}^{sum}(W) \!=\! \max [H(X_1)\!+\!I(V;X_2)\!-\!I(X_1;V)]
\end{equation}
such that $Ew \leq W$, $V \ra X \ra (X_1,X_2)$, and $X_2 \ra V \ra X$. Then, $\bar{p}(x_2|x)$ and $\{p^*(v),p^*(x|v)\}$ and $X=(X_1,\h{X}_2)$ induce the following joint distribution:
\begin{align}
p^*(v,x_1,\h{x}_2,x_2) =p^*(v)p^*(x_1,\h{x}_2|v)\bar{p}(x_2|x_1,\h{x}_2).
\end{align}
Let $p^*(x_1,x_2)$, $p^*(v|x_2)$ and $p^*(\h{x}_2|x_1,v)$ be the corresponding marginals. If $p^*(v,x_1,\h{x}_2,x_2)$ satisfies $X_1 \ra X_2 \ra V$, then $\exists$ a dual BYP with correlated messages $(W_1,W_2)$, which can be associated with a graph $G(\Delta_1, \Delta_2, \Delta_1',\Delta_2', \mu)$, for the source $(X_1,X_2) \sim p^*(x_1,x_2)$ with alphabets $\mc{X}_i$ for $i=1,2$ and a reconstruction alphabet $\mc{\h{X}}_2$, a distortion measure $d: \mc{X}_1 \times \mc{X}_2 \times \mc{\h{X}}_2 \ra \mb{R}^{+}$, and a distortion constraint $D$ such that:
\begin{itemize}
  \item $R_{SBC}^{sum}(W)=R_{BYP}^{sum}(D)$, i.e., $\max [H(X_1)\!+\!I(V;X_2)\!-\!I(X_1;V)]= \min [H(X_1)\!+\!I(V;X_2)\!-\!I(X_1;V)]$ where the maximum is taken over $p(v)$ and $p(x|v)$ with the fixed channel conditional distribution $\bar{p}(x_2|x)$ and $x_1=f(x)$ such that $Ew \leq W$, $V \ra X \ra (X_1,X_2)$, and $X_2 \ra V \ra X$; and the minimum is taken over $p(v|x_2)$, $p(\h{x}_2|x_1,v)$ and the fixed source distribution $p^*(x_1,x_2)$ such that $Ed \leq D$, $X_1 \ra X_2 \ra V$ and $X_2 \ra V \ra (X_1,\h{X}_2)$;
  \item the distributions $p^*(v|x_2)$ and $p^*(\h{x}_2|x_1,v)$ induced from the SBC problem achieve the minimum in the dual BYP,
  \item the correlation structure of the messages $(W_1,W_2)$ of the dual BYP is the same as that of the given SBC problem, i.e, the graph $G(\Delta_1, \Delta_2, \Delta_1',\Delta_2', \mu)$ of the dual BYP is the same as that of the given SBC
\end{itemize}
provided the distortion measure and the distortion constraint are chosen such that
\begin{equation}
  d(x_1,x_2,\h{x}_2)\!=\!-c_2 \log \bar{p}(x_2|x) \!+\! d_0(x_1,x_2)
\end{equation}
and $D=E_{p^*(x_1,x_2) p^*(\h{x}_2|x_1,x_2)}[d(X_1,X_2,\h{X}_2)]$ where $c_2 >0$ and $d_0(x_1,x_2)$ are arbitrary.
\end{thm}

\emph{Proof:} Theorem \ref{thm:main1} can be proved by applying the similar technique used in the proof of Theorem 1 in \cite{pradhan-ram06}. For the sake of brevity, we omit the redundant part of the proof.  The different part of the proof is to show that the correlation structure of the messages in the dual SBC problem is the same as that of the given BYP, and vice versa.

For the given BYP with correlated messages, the correlation structure is determined by the random bipartite graph $\Bbb{G}$ generated from the bin indices of the random codebooks $\Bbb{C}_1$ and $\Bbb{C}_2$ and the jointly typicality of auxiliary random variables $X_1$ and $V$ as shown in Section \ref{sec:ach-proof}. The correlation structure of the dual SBC problem with correlated messages is also determined in a similar way which is shown in the proof of Theorem 1 in \cite{choi-pradhan08}. More precisely, if we choose $U=X_1$ and assume that there is no common message $W_0$ in the proof of Theorem 1 in \cite{choi-pradhan08}, then the correlation structure of the dual SBC problem with correlated messages is also determined by the random graph $\Bbb{G}$ which is generated from the bin indices of the random codebooks and the jointly typicality of auxiliary random variables $X_1$ and $V$.
Note that the same auxiliary random variables $X_1$ and $V$ are used for both BYP and its dual SBC. Therefore, it is obvious that the correlation structure of the messages $(W_1,W_2)$ of the dual SBC problem can be the same as that of the given BYP if the same random codebooks are used for both cases. $\hfill$ $\blacksquare$

\begin{rem}
  Theorem \ref{thm:main1} is similar to Lemma 4 in \cite{stankovic-cheng-xiong06}, presenting the functional duality between BYP and SBC with \emph{independent} messages. Theorem \ref{thm:main1} above is also analogous to the Theorem 1 in \cite{pradhan-ram06}, showing the functional duality between distributed source coding and broadcast channel coding with \emph{independent} messages.

  However, note that there are significant differences as follows; 1) Lemma 4 in \cite{stankovic-cheng-xiong06} and Theorem 1 in \cite{pradhan-ram06} show the functional duality in the case of \emph{independent} messages only, whereas Theorem \ref{thm:main1} above extends the functional duality to the case of \emph{correlated} messages. 2) Moreover, Theorem \ref{thm:main1} specifies the correlation structure of the messages in the two dual problems, while there is no such consideration in \cite{stankovic-cheng-xiong06} and \cite{pradhan-ram06}.
\end{rem}

\section{Conclusion}\label{sec:conclusion}
We have considered a distributed (or multiterminal) source coding problem where two non-communicating encoders represent a pair of correlated sources (for transmission over multiple-access channels) into correlated messages, which can be associated with an undirected nearly semi-regular bipartite graph, and a joint decoder reconstruct the original sources. Here, the reconstruction of one source is lossless and that of the other is lossy with certain distortion criterion.
As a result, we have shown that the given correlated sources can be represented into such graphs with satisfying distortion criterion, providing the information-theoretic achievable rate-distortion region for this problem. Therefore, by merging the results of this paper,  \cite{pradhan-choi-ram07} and \cite{choi09}, we can conclude that a nearly semi-regular bipartite graph can be used as discrete interface in Shannon-style modular approach
for transmission of any (either discrete or continuous) set of correlated sources over the multiple-access channels.

We have also shown that under certain conditions there exists functional duality between our problem, ``Berger-Yeung problem with correlated messages'', and semi-deterministic broadcast channel with correlated messages. We have also specified the correlation structure of two dual problems and the source distortion measure and the channel cost measures for the duality.


\appendix



\section{A characterization of $\e_1(\e)$}\label{app:e1}

\emph{A characterization of $\e_1(\e)$}:
For a precise characterization of the error events, we need a function of
$\epsilon$, and certain properties of typical sets. For any
pair $(U,V)$ of finite-valued random variables, there
exists \cite{cover-thomas,ck} a continuous positive function
$\epsilon_1(\epsilon)$ such that (a) $\epsilon_1(\epsilon) \rightarrow 0$ as $\epsilon
\rightarrow 0$ and (b) for all $\epsilon>0$ (sufficiently small),
there exists an integer $N_0(\epsilon)>0$ such that $\forall
n>N_0(\epsilon)$ the following conditions hold simultaneously:
\begin{align}
2^{n(H(U) -\epsilon_1)} \leq &|\Ae(U)| \leq 2^{n(H(U)+\epsilon_1)}, \label{eqU} \\
2^{n(H(V) -\epsilon_1)} \leq &|\Ae(V)| \leq 2^{n(H(V)+\epsilon_1)}, \label{eqV} \\
2^{n(H(U,V) -\epsilon_1)} \leq &|\Ae(U,V)| \leq 2^{n(H(U,V)+\epsilon_1)}, \label{eqUV}
\end{align}

$\forall u^n \in \Ae(U)$,
\beq
\left|\frac{1}{n} \log \frac{1}{P\{(u^n,V^n)\in \Ae(U,V) \}} - I(U;V) \right| \leq 3\e_1.
\eeq

\bibliographystyle{IEEEtran}

\bibliography{DSC_with_one_distortion_criterion_and_correlated_messages}

\end{document}